# How robust is quicksort average complexity?

Suman Kumar Sourabh[a]

Soubhik Chakraborty[b]*

[a]University Department of Statistics
And Computer Applications
T. M. Bhagalpur University
Bhagalpur - 812007
India

[b]Department of Applied Mathematics
B. I. T. Mesra
Ranchi-835215
India

**Abstract**

The paper questions the robustness of average case time complexity of the fast and popular quicksort algorithm. Among the six standard probability distributions examined in the paper, only continuous uniform, exponential and standard normal are supporting it whereas the others are supporting the worst case complexity measure. To the question -why are we getting the worst case complexity measure each time the average case measure is discredited? -- one logical answer is average case complexity under the universal distribution equals worst case complexity. This answer, which is hard to challenge, however gives no idea as to which of the standard probability distributions come under the umbrella of universality. The morale is that average case complexity measures, in cases where they are different from those in worst case, should be deemed as robust provided only they get the support from at least the standard probability distributions, both discrete and continuous. Regretfully, this is not the case with quicksort.

**\*Corresponding author.**

**Email addresses: soubhikc@yahoo.co.in (S. Chakraborty)**

                **sourabh.suman@rediffmail.com (S. K. Sourabh)**

# 1. Introduction:

It has been clearly stated in several papers ([1]-[4]) and reviews (see e.g. [5]) that

- (i) time is an operation weight and not an operation count [1] [2]

- (ii) a statistical complexity bound weighs rather than counts the operations unlike a mathematical bound. Also, whereas a mathematical bound is operation specific, a statistical bound takes all operations collectively [3]. A table of differences between math bounds and stat bounds in complexity can be found in [3] and [4].

- (iii) It makes sense to work directly on the running time of a program to estimate a statistical bound over a finite and feasible range [2]. This estimate is called empirical O and is written as O with a subscript emp. Of course, we can also estimate a mathematical bound experimentally but in that case the estimate should be count based and operation specific (see [6] for example).

- (iv) although statistical bounds were initially created to make make average complexity a better science [2] [4], they are also useful in giving a certificate on the level of conservativeness of the guarantee giving mathematical bounds in worst case [3]. In this way, worst case complexity, an acknowledged strong area of theoretical computer science, can be made more meaningful. Finally, statistical bounds can easily nullify a tall mathematical claim in best case as in [3].

- (v) The credibility of the bound-estimate depends on proper design and analysis of a special kind of computer experiment whose response is a complexity, rather than output, such as time [4]. Thus for example the output in a sorting algorithm is the sorted array. But in the computer experiments involved in the present work, the response is time.

- (vi) A computer experiment can be run only over a finite range. Therefore the finite range concept is important to set up a link between research in computer experiments with that in algorithmic complexity. See also [7] and the relevant references cited therein in addition to our works.

- (vii) Ref. [8] gives further insight into statistical bounds and shows that such bounds can be both probabilistic [9] and non-probabilistic.

The paper makes use of most of these concepts and questions the robustness O(nlogn) average case time complexity of Hoare's fast and popular Quicksort algorithm. An excellent reference on the historical perspectives of this algorithm, with special emphasis on several improvements tried by different authors, can be found in [10] which also gives an interesting empirical comparison of these improved versions. This includes removal of interchanges achieved by two members of our research team [11].

For the benefit of the reader, an appendix to this paper gives the codes used.

## 2. Empirical Results:

This section provides a number of interesting empirical results on the fast and popular Quicksort algorithm and questions the robustness of average complexity measure of this algorithm, namely O(nlogn), derived assuming uniform distribution, for non-uniform inputs (both discrete and continuous case). [The base 2 of the logarithm has no effect on the O notation and hence not considered].

.The observed average time (in sec.) of 10-trials for sorting different discrete and continuous distribution inputs of sample size *n*. The observations are taken on fixed parameters for different distributions, but with varying sample size *n* in the range [5000, 50000].

The following observations are taken on the system whose specifications are given below:

**System Specifications**

Processor        Intel Pentium ® 4 CPU 3.0 GHz
Hard Disk        160 GB
RAM              448 MB
Operating System    Windows XP Professional
                 Version 2002
                 Service Pack 2

The observed average times (in sec.) for *Quicksort* are depicted in Table – 1.

**Table 1: Table of mean sorting time in seconds for different distribution inputs for Quicksort**

| n | nlogn | Binomial (m=100, p=0.5) | Poisson $\lambda$ =1 | Discrete Uniform [1,2,...,k] k=50 | Continuous Uniform [0,1] | Exponential [mean = 1] | Standard Normal (0,1) |
|---|---|---|---|---|---|---|---|
| 5000 | 18494.85 | 0.0047 | 0.0047 | 0.0015 | 0.0016 | 0.0016 | 0.0031 |
| 10000 | 40000.00 | 0.0095 | 0.0172 | 0.0031 | 0.0031 | 0.0047 | 0.0063 |
| 15000 | 62641.37 | 0.0091 | 0.0422 | 0.0062 | 0.0062 | 0.0078 | 0.0062 |
| 20000 | 86020.60 | 0.0156 | 0.0719 | 0.0062 | 0.0062 | 0.0109 | 0.0110 |
| 25000 | 109948.50 | 0.0266 | 0.1140 | 0.0093 | 0.0093 | 0.0110 | 0.0109 |
| 30000 | 134313.64 | 0.0345 | 0.1609 | 0.0156 | 0.0157 | 0.0156 | 0.0140 |
| 35000 | 159042.38 | 0.0421 | 0.2188 | 0.0203 | 0.0156 | 0.0156 | 0.0154 |
| 40000 | 184082.40 | 0.0579 | 0.2812 | 0.0218 | 0.0157 | 0.0171 | 0.0189 |
| 45000 | 209394.56 | 0.0735 | 0.3625 | 0.0282 | 0.0204 | 0.0202 | 0.0219 |
| 50000 | 234948.50 | 0.0844 | 0.4453 | 0.0391 | 0.0235 | 0.0219 | 0.0233 |

**Table 2: Table of standard deviation of sorting time in seconds for different distribution inputs for Quicksort**

| n | Binomial (m=100, p=0.5) | Poisson $\lambda$ =1 | Discrete Uniform [1,2,...,k] k=50 | Continuous Uniform [0,1] | Exponential [mean = 1] | Standard Normal (0,1) |
|---|---|---|---|---|---|---|
| 5000 | 0.007573 | 0.007573 | 0.004743 | 0.005060 | 0.005060 | 0.006540 |
| 10000 | 0.008182 | 0.005224 | 0.006540 | 0.006540 | 0.007573 | 0.008138 |
| 15000 | 0.007838 | 0.007052 | 0.008011 | 0.008011 | 0.008230 | 0.008011 |
| 20000 | 0.000516 | 0.008103 | 0.008011 | 0.008011 | 0.007534 | 0.007601 |
| 25000 | 0.007560 | 0.007601 | 0.008015 | 0.008015 | 0.007601 | 0.007534 |
| 30000 | 0.006604 | 0.007666 | 0.000516 | 0.000483 | 0.000516 | 0.004944 |
| 35000 | 0.007445 | 0.007315 | 0.007861 | 0.000516 | 0.000516 | 0.000516 |
| 40000 | 0.007534 | 0.000422 | 0.008364 | 0.000483 | 0.005259 | 0.006919 |
| 45000 | 0.007487 | 0.006604 | 0.006443 | 0.007792 | 0.007927 | 0.008062 |
| 50000 | 0.008058 | 0.019833 | 0.008333 | 0.008127 | 0.008062 | 0.008125 |

Based on table 1, we compared empirical models corresponding to O(nlogn) and O($n^2$) complexity for each distribution input separately. Our results are summarized in sub sections 2.1-2.6.

## 2.1. Average Case Complexity for Binomial Distribution Inputs

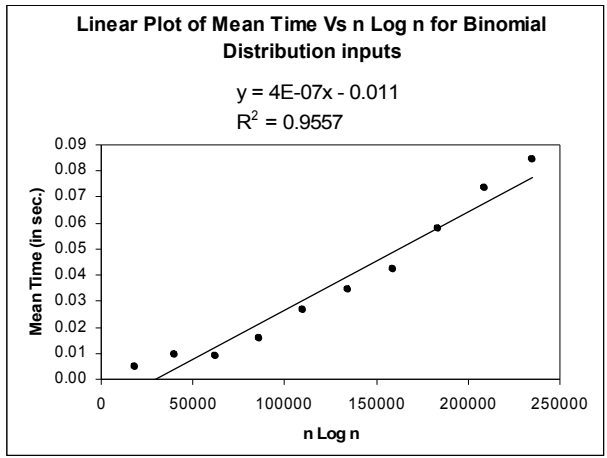

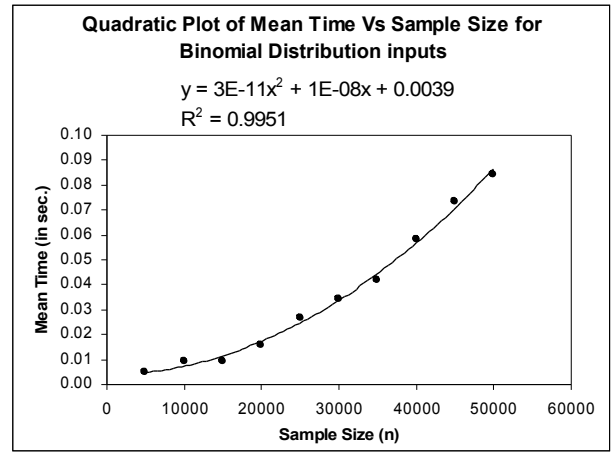

Fig. 1                                   Fig. 2

Experimental results as shown in fig. 1 and 2 are not supporting $O(n \log n)$ complexity; rather they are supporting $O(n^2)$ complexity for Binomial distribution inputs.

We write $y_{avg}(n) = O_{emp}(n^2)$. Explanation for such contradictions is given in the concluding section (sec 3). Other issues are also discussed.

## 2.2 Average Case Complexity for Poisson Distribution Inputs

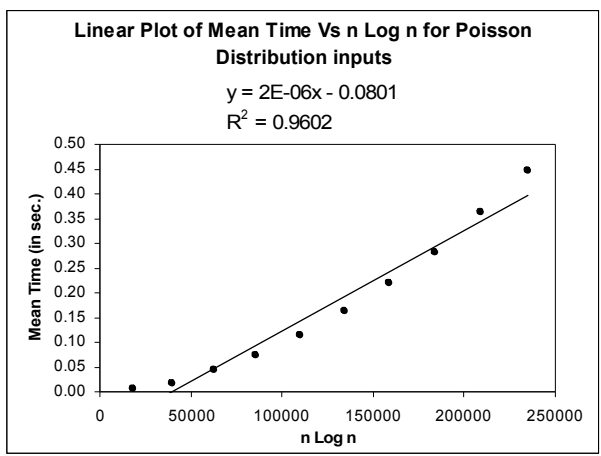

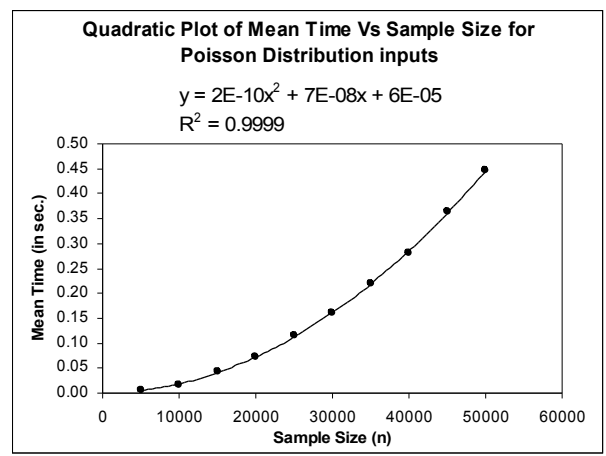

Fig. 3                                   Fig. 4

Experimental results as shown in fig. 3 and 4 are not supporting $O(n \log n)$ complexity rather they are supporting $O(n^2)$ complexity. Best support for $O(n^2)$ complexity was found for Poisson distribution inputs. We write $y_{avg}(n) = O_{emp}(n^2)$

## 2.3 Average Case Complexity for Discrete Uniform Distribution Inputs

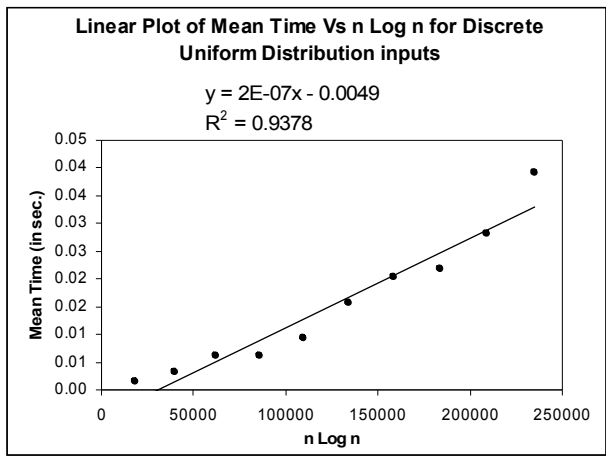
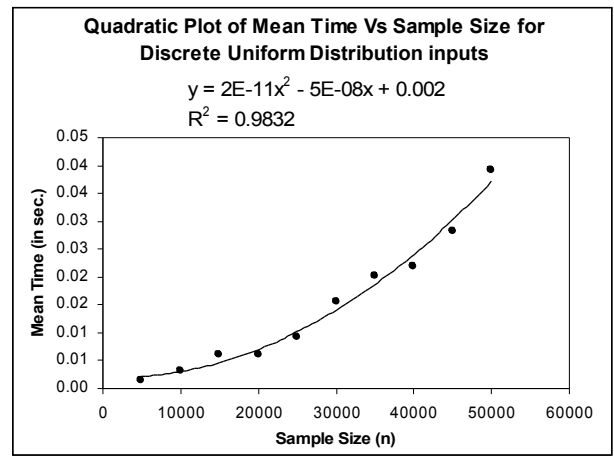

Fig. 5                                      Fig. 6

Experimental results as shown in fig. 5 and 6 are again supporting $O(n \log n)$ complexity less than they are supporting $O(n^2)$ complexity for Discrete Uniform distribution inputs.

We write $y_{avg}(n) = O_{emp}(n^2)$

## 2.4 Average Case Complexity for Continuous Uniform Distribution Inputs

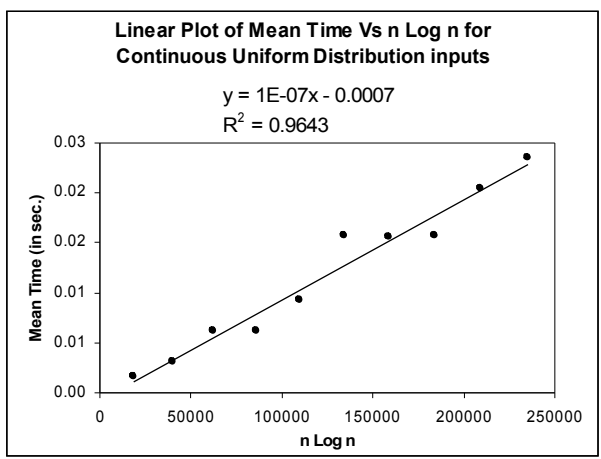
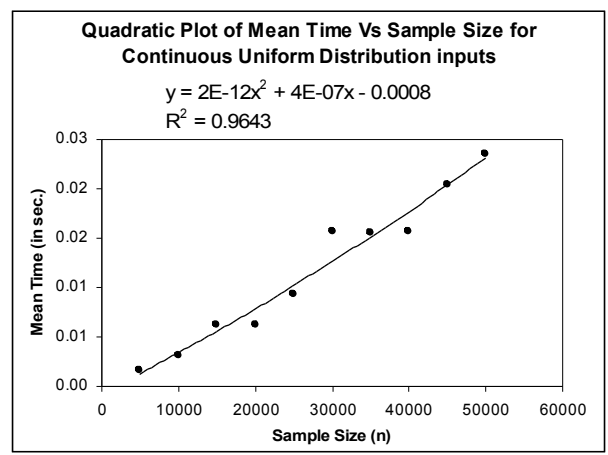

Fig. 7                                      Fig. 8

Experimental results as shown in fig. 7 and 8 are supporting $O(n \log n)$ complexity and they are not supporting $O(n^2)$ complexity any better for continuous Uniform distribution inputs. Best results are obtained confirming the theory here only.

We write $y_{avg}(n) = O_{emp}(n \log n)$

## 2.5 Average Case Complexity for Exponential Distribution Inputs

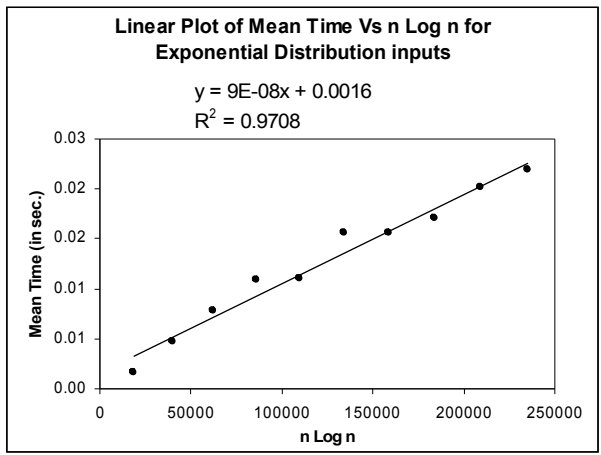

Fig. 9

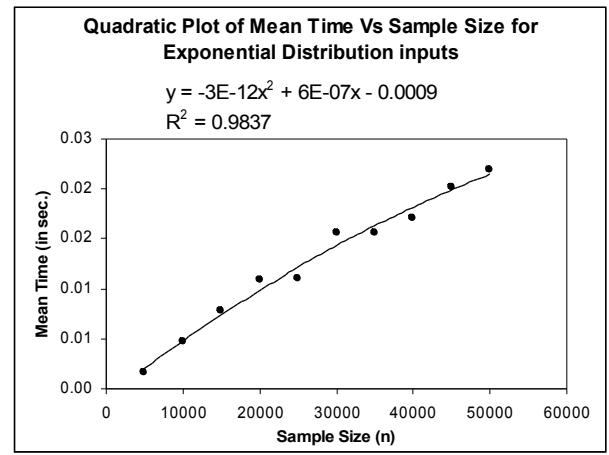

Fig. 10

Experimental results as shown in fig. 9 and 10 are supporting $O(n \log n)$ complexity and they are not supporting $O(n^2)$ complexity any better for Exponential distribution inputs.

We write $y_{avg}(n) = O_{emp}(nlogn)$

## 2.6 Average Case Complexity for Standard Normal Distribution Inputs

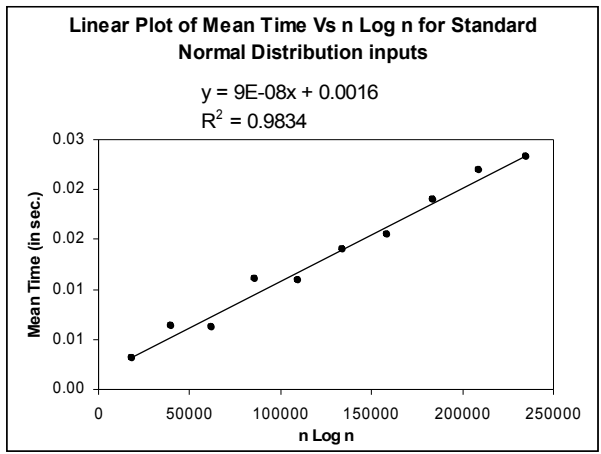

Fig. 11

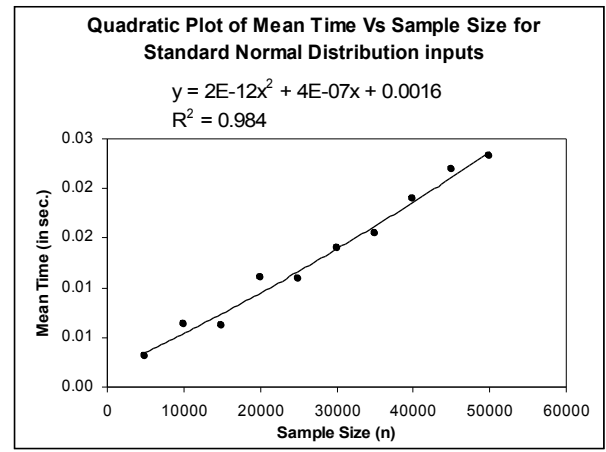

Fig. 12

Experimental results as shown in fig. 11 and 12 are supporting $O(n \log n)$ complexity and they are not supporting $O(n^2)$ complexity any better for Standard Normal input.

We write $y_{avg}(n) = O_{emp}(nlogn)$

## 3. Conclusion

Among the six standard probability distributions examined in the paper, only continuous uniform, exponential and standard normal are supporting the O(nlogn) complexity whereas Binomial, Poisson and discrete uniform are supporting the O($n^2$) complexity. To the question "Why are we getting the worst case complexity measure each time the average case measure is discredited?" one logical answer is "average case complexity under the universal distribution equals worst case complexity"[12]. However, this brilliant paper gives no idea as to which of the standard probability distributions come under the umbrella of "universality". The morale is that average case complexity measures, in cases where they are different from those in worst case, are robust provided only *they get the support from at least the standard probability distributions, both discrete and continuous*. Regretfully, this is not the case with Hoare's Quicksort.

Our investigatioins are ongoing as to whether the lack of support of the O(nlogn) complexity for discrete distributions is due to the presence of ties (given that the probability of a tie is zero in inputs from continuous distributions). See also [13] where Knuth's proof came "under a cloud" on a similar ground having to do with ties although that was not a paper where we worked on time or weight in any sense whatsoever. Another difference is that in [13], the "clouds" were caused by the absence of ties wheras here it is their presence that is of interest. Since ties are crucial in complexity analysis, we were not satisfied with our maiden dip into the "gold standard" [14] in a paper where ties were involved and recently made a second dip [15]. Hopefully there will be many more dips, more "clouds" and some "rains"** too!

As a final comment**,** since comparisons dominate a sorting algorithm, the reader is encouraged to cross check the present results against mean comparisons experimentally counted rather than working on time directly. In a complex code such as one in partial differential equation, it is hard to guess the pivotal operation for taking the expectation. See [2] on how the dominance of multiplication can be challenged by another dominant operation (comparison) in matrix multiplication. The method and ideas given in the present paper have a novelty and are very general no matter how complex the code is. This point must be understood. To the question why we took only 10 trials (table 1) and not 500 (say) for each n, the simple answer is that the standard deviations are small enough (table 2) so that the mean of 10 trials is not expected to differ by much from the mean of 500 trials. *But this won't be the case if we work on comparisons instead (verify!)* and 500 trials (say) ought to be taken at each point of n to get a reliable measure of mean, accounting for much valuable time of the researcher. This gives the reader a second strong motivation to work directly on time.

[Concluded]

---------------------------------------------------------------------------------------------------------------

**breakthrough.

# APPENDIX

## Function : quick Sort

```
// Quick Sort Function

void partition(int *x, int lb, int ub, int &pj)
    {
    int a, down, up, temp;
    a=x[lb];
    up=ub;
    down=lb;
    while(down < up)
      {
       while(x[down] <= a && down < ub)
          down++;
       while(x[up] > a)
           up--;
       if(down < up)
          {
           temp=x[down];
           x[down]=x[up];
           x[up]=temp;
          }
       }
       x[lb]=x[up];
       x[up]=a;
       pj=up;
    }

void quicksort(int *x, int lb, int ub)
    {
    int j=1;
    if(lb > ub)
      return;
    else
          {
           partition(x,lb,ub,j);
           quicksort(x,lb,j-1);
           quicksort(x,j+1,ub);
          }
    }
```

**Program 1 : Average Complexity for Binomial Distribution Inputs**

```
/***********************************************************/
/*  QUICK SORT ELAPSED TIME (in sec) IN SORTING SAMPLE OF  */
/*         SIZE N FOR BINOMIAL DISTRIBUTION INPUTS         */
/***********************************************************/

#include <stdlib.h>
#include <iostream.h>
#include <conio.h>
#include <sys/timeb.h>
#include <time.h>

void partition(int *x, int lb, int ub, int &pj);

void quicksort(int *x, int lb, int ub);
```

```
void main()
{
 int n,*a,m,s;
 float p,r;
 clock_t start, end;

 clrscr();

 cin>>n;
 cin>>m;
 cin>>p;

 a=new int[n];

 randomize();

 for(int i=0;i<n;i++)
      {
       s=0;
       for(int j=0;j<m;j++)
            {
             r=(float)rand()/RAND_MAX;
             if(r<p)
                  ++s;
            }
        *(a+i)=s;
      }

 start=clock();
      quicksort(a,0,n-1);
 end=clock();

 cout.precision(4);
 cout.setf(ios::showpoint);
 cout<<endl<<(end-start)/CLK_TCK;
}
```

**Program 2 : Average Complexity for Poisson Distribution Inputs**

```
/***********************************************************/
/*  QUICK SORT ELAPSED TIME (in sec) IN SORTING SAMPLE OF   */
/*          SIZE N FOR POISSON DISTRIBUTION INPUTS          */
/***********************************************************/

#include <stdlib.h>
#include <iostream.h>
#include <conio.h>
#include <sys/timeb.h>
#include <time.h>

void partition(int *x, int lb, int ub, int &pj);
void quicksort(int *x, int lb, int ub);

void main()
{
 int n,*a,x;
 float Lambda, p, b=(float) exp((-1)*Lambda),r;
 clock_t start, end;

 clrscr();
```

```
 randomize();

 cin>>Lambda;
 cin>>n;

 a=new int[n];

 for(int i=0;i<n;i++)
     {
       p=1.0;
       for(int j=1;j<5000;j++)
           {
            r=(float)rand()/RAND_MAX;
            p=p*r;
            if(p<b)
                {
                  x=j-1;
                  break;
                }
           }
       *(a+i)=x;
     }

 start=clock();
      quicksort(a,0,n-1);
 end=clock();

 cout.precision(4);
 cout.setf(ios::showpoint);
 cout<<endl<<(end-start)/CLK_TCK;
 delete a;
}
```

**Program 3 : Average Complexity for Discrete Uniform Distribution Inputs**

```
/*************************************************************/
/*   QUICK SORT ELAPSED TIME (in sec) IN SORTING SAMPLE OF   */
/*      SIZE N FOR DISCRETE UNIFORM DISTRIBUTION INPUTS      */
/*************************************************************/

#include <stdlib.h>
#include <iostream.h>
#include <conio.h>
#include <sys/timeb.h>
#include <time.h>

void partition(int *x, int lb, int ub, int &pj);

void quicksort(int *x, int lb, int ub);

void main()
{
 int n,*a,k;
 float r;
 clock_t start, end;

 clrscr();
 randomize();

 cin>>k;
 cin>>n;
```

```
 a=new int[n];

 for(int i=0;i<n;i++)
      {
       r=(float)rand()/RAND_MAX;
       *(a+i)=(int)(k*r)+1;
      }

 start=clock();
     quicksort(a,0,n-1);
 end=clock();

 cout.precision(4);
 cout.setf(ios::showpoint);
 cout<<endl<<(end-start)/CLK_TCK;
 delete a;
}
```

**Program 4 : Average Complexity for Continuous Uniform Distribution Inputs**

```
/*************************************************************/
/*  QUICK SORT ELAPSED TIME (in sec) IN SORTING SAMPLE OF    */
/*    SIZE N FOR CONTINUOUS UNIFORM DISTRIBUTION INPUTS      */
/*************************************************************/

#include <stdlib.h>
#include <iostream.h>
#include <conio.h>
#include <sys/timeb.h>
#include <time.h>

void partition(double *x, int lb, int ub, int &pj);

void quicksort(double *x, int lb, int ub);

void main()
{
 int n;
 int theta;
 double *a;
 clock_t start, end;

 clrscr();
 randomize();

 cin>>theta;
 cin>>n;

 a=new double[n];

 for(int i=0;i<n;i++)
       *(a+i)= (double)rand()/RAND_MAX*theta;

 start=clock();
     quicksort(a,0,n-1);
 end=clock();

 cout.precision(4);
 cout.setf(ios::showpoint);
```

```
    cout<<endl<<(end-start)/CLK_TCK;
     delete a;
    }
```

**Program 5 : Average Complexity for Exponential Distribution Inputs**

```
/***********************************************************/
/*   QUICK SORT ELAPSED TIME (in sec) IN SORTING SAMPLE OF   */
/*         SIZE N FOR EXPONENTIAL DISTRIBUTION INPUTS        */
/***********************************************************/

#include <stdlib.h>
#include <iostream.h>
#include <conio.h>
#include <sys/timeb.h>
#include <time.h>

void partition(double *x, int lb, int ub, int &pj);

void quicksort(double *x, int lb, int ub);

void main()
{
 int n,theta;
 double *a;
 double r;
 clock_t start, end;

 clrscr();
 randomize();

 cin>>theta;
 cin>>n;

 a=new double[n];

 for(int i=0;i<n;i++)
       {
        r=(double) rand()/RAND_MAX;
           if(r<=0)
            {
             --i;
             continue;
            }
         *(a+i)=(double)(-1)/theta*log(r);
       }
 start=clock();
       quicksort(a,0,n-1);
 end=clock();

 cout.precision(4);
 cout.setf(ios::showpoint);
 cout<<endl<<(end-start)/CLK_TCK;
 delete a;
}
```

**Program 6 : Average Complexity for Standard Normal Distribution Inputs**

```c
/***********************************************************/
/*  QUICK SORT ELAPSED TIME (in sec) IN SORTING SAMPLE OF   */
/*      SIZE N FOR STANDARD NORMAL DISTRIBUTION INPUTS      */
/***********************************************************/

#include <stdlib.h>
#include <iostream.h>
#include <conio.h>
#include <sys/timeb.h>
#include <time.h>

void partition(double *x, int lb, int ub, int &pj);
void quicksort(double *x, int lb, int ub);

void main()
{
 int n, mean, var;
 double *a, u1, u2, z1, z2;
 clock_t start, end;

 clrscr();
 randomize();

 cin>>var;
 cin>>mean;
 cin>>n;

 a=new double[n];

 for(int i=0;i<n/2;i++)
     {
      u1=(double) rand()/RAND_MAX;
      u2=(double) rand()/RAND_MAX;
         if(u1<=0)
         {
          --i;
          continue;
         }
      z1 = sqrt((-2)*log(u1)) * cos(8*atan(1)*u2);
      z2 = sqrt((-2)*log(u1)) * sin(8*atan(1)*u2);
      *(a+i) = z1*mean*var;
      *(a+n/2+i) = z2*mean*var;
     }
 start=clock();
     quicksort(a,0,n-1);
 end=clock();

 cout.precision(2);
 cout.setf(ios::showpoint);
 cout<<(end-start)/CLK_TCK<<endl;
 delete a;
}
```